\documentclass{ws-procs9x6}

\newcommand{\ie}{{\it i.e.}}
\newcommand{\eg}{{\it e.g.}}
\newcommand{\cf}{{\it cf.}}

\newcommand{\gev}{{\rm GeV}}

\newcommand{\qu}{{\rm q}}
\newcommand{\qb}{${\rm\bar q}$}
\newcommand{\qbm}{{\rm\bar q}}
\newcommand{\qq}{\qu\qb\ }

\newcommand{\bs}[1]{\boldsymbol{#1}}

\newcommand{\lqcd}{\Lambda_{QCD}}
\newcommand{\as}{\alpha_s}

\newcommand{\M}{{\cal M}}
\newcommand{\gsim}{\buildrel > \over {_\sim}}
\newcommand{\lsim}{\buildrel < \over {_\sim}}

\newcommand{\order}[1]{${\cal O}\left(#1 \right)$}
\newcommand{\morder}[1]{{\cal O}\left(#1 \right)}
\newcommand{\eq}[1]{(\ref{#1})}

\newcommand{\halft}{{\textstyle \frac{1}{2}}}
\newcommand{\beq}{\begin{equation}}
\newcommand{\eeq}{\end{equation}}

\newcommand{\beqa}{\begin{eqnarray}}
\newcommand{\eeqa}{\end{eqnarray}}

\newcommand{\inv}[1]{\frac{1}{#1}}

\begin{document}

\hfill HIP-2007-43/TH

\title{INCLUSIVE PERSPECTIVES\footnote{Concluding talk at the Workshop on {\it Exclusive Reactions at High Momentum Transfer}, Jefferson Lab, 21-24 May 2007.}}

\author{PAUL HOYER}

\address{Department of Physical Sciences, University of Helsinki,\\
P.O.B. 64, FIN-00014 University of Helsinki, Finland\\
www.helsinki.fi/\~{}hoyer}

\begin{abstract}
I discuss the relation between inclusive and exclusive dynamics suggested by Bloom-Gilman duality. Duality implies the simultaneous applicability of two distinct limits, the standard DIS limit of hard inclusive processes taken at fixed $x_B$ and a limit where the hadronic mass is held fixed. I review experimental evidence for the relevance of the fixed mass limit in inclusive processes at high $x_F$. Semi-local duality suggests that inclusive and exclusive processes occur on the same target Fock states. DIS scaling then implies that the Fock states contributing to hard exclusive processes have a large transverse size, \ie, that the hard scattering occurs off a single parton which carries a large fraction of the hadron momentum.
\end{abstract}


\bodymatter


\subsection*{The inclusive-exclusive connection} I shall take an inclusive perspective on exclusive dynamics, which is the focus of this workshop. Following Drell, Yan and West \cite{Drell:1969km} we expect, as illustrated in Fig.~1, to recover exclusive form factors from Deep Inelastic Scattering (DIS) in the limit of $x_B\to 1$, where $x_B = Q^2/2m\nu$ is the Bjorken variable.

\begin{figure}[h]
\begin{center}
\psfig{file=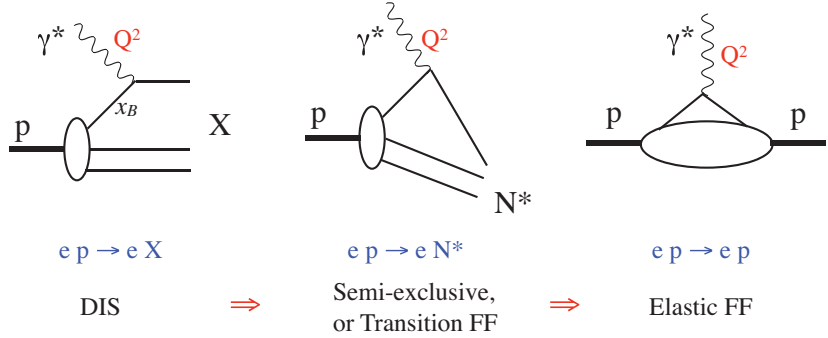,width=11cm}
\end{center}
\caption{The inclusive -- exclusive connection in $ep\to eX$. As $x_B = Q^2/2m\nu \to 1$ the mass $W$ of the hadronic final state $X$ decreases towards the target nucleon mass. There is a continuous connection between inclusive DIS, $p\to N^*$ transition form factors and the nucleon elastic form factor.}
\label{fig1}
\end{figure}

As first observed by Bloom and Gilman \cite{Bloom:1970xb}, and more recently confirmed also for nuclear target and spin dependence by data from Jefferson Lab and DESY \cite{Melnitchouk:2005zr}, the inclusive-exclusive connection works much better than anyone could have expected. In Fig.~2 the $ep\to eN^*$ cross sections for $N^*=\Delta(1232)$ and $S_{11}(1535)$, which determine the corresponding $p\to N^*$ transition form factors, are compared to the scaling (high $Q^2$) DIS cross section at common values of the Nachtmann variable $\xi \simeq x_B$,
\beq\label{xidef}
\xi = \frac{2x_B}{1+\sqrt{1+4x_B^2 m^2/Q^2}} = x_B\left(1-x_B^2\frac{m^2}{Q^2}+\ldots\right)
\eeq
The $x_B$ (and $\xi$) of a given $N^*$ tends to unity, $x_B \to1$, with increasing virtuality $Q^2$ of the photon,
\beq\label{mrel}
M_{N^*}^2 = m_N^2 + \frac{1-x_B}{x_B}\,Q^2
\eeq
Bloom-Gilman duality refers to the remarkable fact that the $N^*$ cross sections at low $Q^2$ equal the (smooth, scaling) DIS cross section measured at high $Q^2$ but at the same value of $\xi$.

\begin{figure}[h]
\begin{center}
\psfig{file=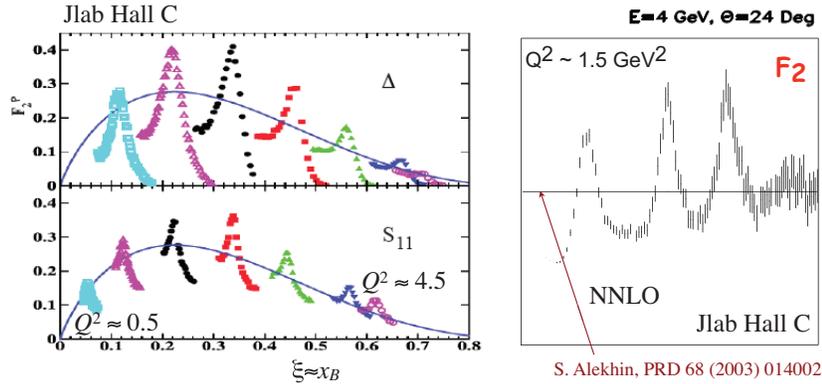,width=11cm}
\end{center}
\caption{Bloom-Gilman duality \cite{Melnitchouk:2005zr}. Left: The proton structure function $F_2(\xi)$ for $ep \to e+X$ is plotted in the $X=\Delta(1232)$ and $X =S_{11}(1535)$ resonance regions for several values of $Q^2$ in the range $0.5 \ldots 4.5\ \gev^2$. The Nachtmann variable $\xi$ defined in \eq{xidef} equals $x_B$ up to target mass corrections. The smooth curve is a fit to the Jlab data at high $Q^2$. Right: The $F_2$ structure function in the resonance region at $Q^2 = 1.5\ \gev^2$ compared to the high $Q^2$ NNLO fit from Ref. \cite{Alekhin:2002fv}.}
\label{fig2}
\end{figure}

The duality between exclusive resonances and inclusive DIS is all the more impressive since exclusive scattering is {\it coherent} on the entire target, whereas DIS scaling is obtained from {\it incoherent} scattering on single partons. According to \eq{mrel} the high $Q^2$ limit of exclusive  processes is taken with $x_B\to 1$:
\beq
Q^2 \to\infty\ {\rm at\ fixed}\ (1-x_B)Q^2 \ \ \ \ \ {\rm (BB\ limit)}\label{bbl}
\eeq
Berger and Brodsky \cite{Berger:1979du} first pointed out novel coherence effects in this BB limit of the Drell-Yan process $\pi N \to \mu^+\mu^- X$\footnote{Berger and Brodsky found multiparton coherence even when $Q(1-x)$ was held fixed. This earlier onset of coherent effects is apparently due to the leading twist quark distribution of the pion, $f_{q/\pi}(x)$, being suppressed for $x\to 1$ due to the helicity mismatch between the pion and the quark.}.
The BB limit \eq{bbl} may be contrasted with the usual Bjorken limit of hard inclusive processes,
\beq
Q^2 \to\infty\ {\rm at\ fixed}\ x_B\hspace{2cm} {\rm (Bj\ limit)}\label{bjl}
\eeq
which is the basis of the twist expansion. The twist expansion does not apply in the BB limit \eq{bbl}, where (as we shall presently discuss) scattering on several partons remains coherent.

The relation between hard inclusive and exclusive processes demonstrated by duality suggests that DIS is not as incoherent as often thought. In fact, it has for some time been realized that even in the Bj limit the hard photon interaction of DIS remains coherent with soft rescattering of the struck quark on spectators in the target \cite{Brodsky:2002ue}. Recently, this has led to doubts about the validity of factorization in hard hadronic processes \cite{Collins:2007nk}.

Conversely, duality shows that hard exclusive scattering has features in common with inclusive dynamics. Unfortunately, little is known about the general properties of scattering in the BB limit \eq{bbl}, \eg, how the hard subprocess of scale $Q$ may be factored from incoherent soft processes in this limit. Higher twist corrections to DIS have been observed to increase at large $x_B$ \cite{Virchaux:1991jc}, and are expected to be governed by the scale $Q^2(1-x_B)$ (which is fixed in the BB limit). The dominant production mechanism of heavy quarks was shown to be multiple soft scattering in the target (rather than hard scattering on a single target parton) when the quark pair mass $M_{Q\bar Q}$ increases with the momentum fraction $x_F$ carried by the quarks such that $(1-x_F)M_{Q\bar Q}^2$ is fixed~\cite{Brodsky:1991dj}.

In the following I shall recall some general features of coherence in the BB limit, consider experimental evidence that this limit is relevant, and argue that it merits further study.

\subsection*{Hard-Soft Coherence in large x Fock States}

Partons are mutually coherent when their lifetimes (inverse energies) are commensurate. Interactions of coherent partons are added at the amplitude level, \ie, interference effects cannot be neglected. The Light-Front (LF) energy\footnote{When the longitudinal momentum components are large compared to masses and transverse momenta coherence measured by the LF and ordinary energies give the same result, since $E=\sqrt{p_\parallel^2+p_\perp^2+m^2}\simeq |p_\parallel|+(p_\perp^2+m^2)/2|p_\parallel|$. The $|p_\parallel|$ terms cancel in energy differences due to momentum conservation.} of a Fock state with total momentum $P$ is
\beq\label{lfen}
P^- \equiv P^0-P^z = \sum_i \frac{p_{i\perp}^2+m_i^2}{x_iP^+}\hspace{1cm} \left(\sum_i x_i=1\right)
\eeq
where $x_i,\ \bs{p}_{i\perp}$ and $m_i$ denote the momentum fraction, transverse momentum and mass of parton $i$, respectively. Contributions to $P^-P^+$ of order $Q^2$ can thus arise in two distinct ways:
\begin{itemize}
\item From {\it hard} partons with $p_{i\perp}^2 \sim Q^2$ or $m_{i}^2 \sim Q^2$
\item From {\it soft} partons with $p_{i\perp}^2 \sim m_{i}^2 \sim \lqcd^2$ but $x_i \sim \lqcd^2/Q^2$
\end{itemize}
Thus hard partons with large $x_i$ can be coherent with soft partons of small $x_i$. If, as in the BB limit, the hard parton takes nearly all the momentum,
\beq\label{xto1lim}
x\to 1\ \ {\rm with} \ \ (1-x)p_\perp^2 \sim \lqcd^2\ \ {\rm fixed}
\eeq
then all the other partons have small $x_i \sim \lqcd^2/p_\perp^2$ and give contributions of \order{p_\perp^2} to $P^-P^+$ in \eq{lfen}. Thus the full Fock state interacts coherently in the BB limit.

Asymmetric Fock states where one quark carries nearly all the momentum are not as unusual as they may seem at first. A good example is provided by DIS itself (Figs.~3 and~4).
%
\begin{figure}[h]
\begin{center}
\psfig{file=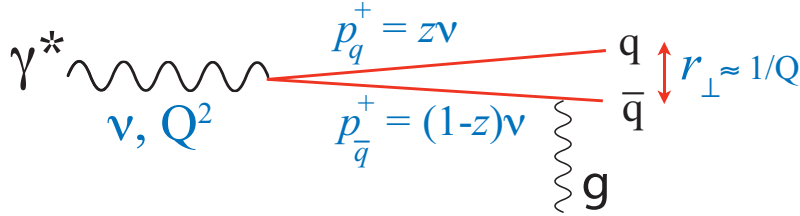,width=11cm}
\end{center}
\caption{DIS as viewed in the target rest frame. The virtual photon fluctuates into a \qq pair which interacts with a target gluon. When the momentum fractions $z$ and $1-z$ of the quarks are similar the pair has a small transverse size $r_\perp \sim 1/Q$ and the gluon is hard.}
\label{fig3}
\end{figure}
%
In the target rest frame, where the virtual photon has large positive longitudinal momentum ($q^+ \simeq 2\nu$), the photon typically fluctuates into a \qq pair before interacting in the target. The momentum fraction $z$ carried by the quark is distributed according to the splitting function
\beq\label{splitfn}
\frac{dP(\gamma\to \qu\qbm)}{dz} \propto \left[z^2+(1-z)^2 \right]
\eeq
The quark and antiquark typically carry commensurate momentum fractions $z = \morder{\halft}$. However, only \qq Fock states which have lifetimes of the same order as the virtual photon are coherent with (and hence can contribute to) the hard process. $P^-P^+ \sim  \morder{Q^2}$ in \eq{lfen} requires $p_{\qu\perp}\sim Q$. Both quarks are `hard' and the size of the Fock state is $r_\perp \sim 1/p_{\qu\perp} \sim 1/Q$. The quark pair is a color singlet and its cross section in the target is $\sigma(\qu\qbm) \sim 1/Q^2$. Color transparency forces the target gluon to be hard and perturbative. The corresponding DIS subprocess $\gamma^* g\to \qu\qbm$ is therefore of higher order in $\as$.

\begin{figure}[h]
\begin{center}
\psfig{file=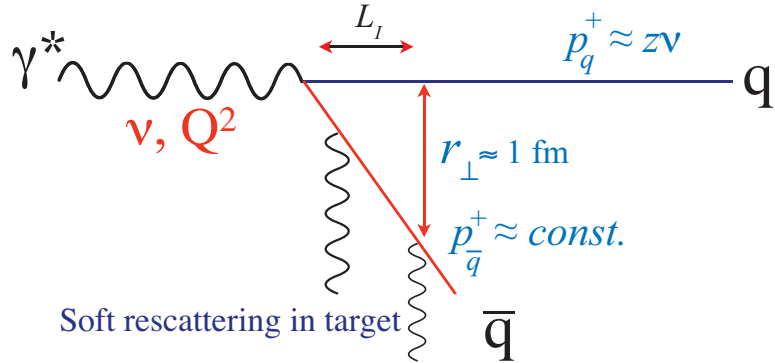,width=11cm}
\end{center}
\caption{If the virtual photon fluctuates into an asymmetric \qq pair, such that the momentum of the antiquark in the target does not grow with $\nu$ and $Q^2$, the pair has a large transverse size $r_\perp \sim 1$ fm. The nonperturbative scattering of the pair in the target (which occurs within the Ioffe length $L_I$ of the photon vertex) determines the DIS cross section.}
\label{fig4}
\end{figure}

The lowest order (parton model) subprocess $\gamma^* q \to q$ appears as an {\it endpoint} contribution where $1-z \sim \lqcd^2/Q^2$ (Fig.~4). The antiquark then has finite momentum in the target\footnote{This is obviously required in order that the antiquark entering the target can equivalently be interpreted as a quark emerging from the target.} even in the Bj limit \eq{bjl}, $p_{\qbm}^+ = (1-z)q^+ \sim \lqcd^2/mx_B$. Because the photon splitting function \eq{splitfn} is finite for $z\to 1$ the probability for the asymmetric splitting is given by the width of the $z$ interval: $P(\gamma\to \qu\qbm) \sim \Delta z \sim \lqcd^2/Q^2$. The asymmetric Fock state is coherent with the virtual photon since the product $P^-P^+$ in \eq{lfen} is of \order{Q^2} due to the `soft' antiquark contribution ($x_{\qbm} \sim \lqcd^2/Q^2$). With $p_{\qu\perp} \sim \lqcd$ the transverse size of the Fock state is large and it interacts in the target with a normal hadronic cross section. The resulting DIS cross section thus scales dimensionally, $\sigma_{DIS} \sim P(\gamma\to \qu\qbm)/\lqcd^2 \sim 1/Q^2$. The gluons in Fig.~4 are nonperturbative and represent the soft scattering of the antiquark in the target.

For our present discussion it is essential to note that the soft interactions of the antiquark are coherent with the virtual photon interaction as long as they occur within the Ioffe length $L_I \simeq 1/2m_Nx_B$. The photon coherence length is Lorentz dilated in the target rest frame,
\beq\label{ioffe}
L_I \simeq \inv{Q}\cdot \frac{\nu}{Q} = \frac{\nu}{Q^2}
\eeq
and thus remains finite in the Bj limit. Soft, coherent interactions of the antiquark are unsuppressed and in fact {\it required} for the fast quark to materialize as a jet in the final state (in $A^+=0$ gauge the fast quark does not interact).

The possibility that soft interactions of partons with small momentum fractions affect the scattering of hard partons with high $x$ may be relevant also for the dynamics of exclusive processes. The magnitude of such effects depends on the behaviour of hadron wave functions in the limit where one parton carries nearly all the hadron momentum. This is a non-perturbative issue which at present must be decided by experiment. I next review some data involving hadron wave functions which points to the relevance of these coherence effects.

\subsection*{Experimental hints}

\subsubsection*{Lepton pair production}

\begin{figure}[h]
\begin{center}
\psfig{file=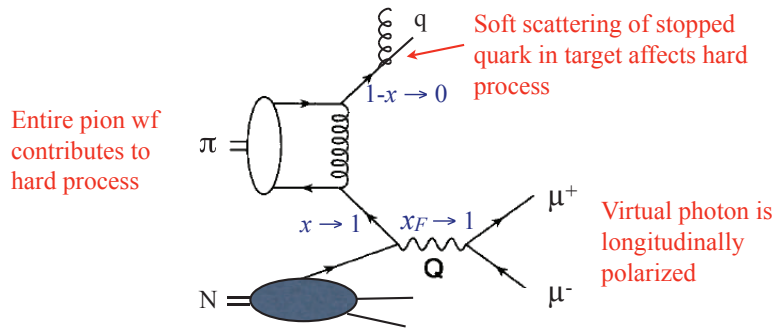,width=11cm}
\end{center}
\caption{The Drell-Yan process $\pi N \to \mu^+\mu^- X$ in the BB limit \eq{bbl} where the momentum fraction of the lepton pair  $x_F \to 1$ such that $(1-x_F)Q^2$ is fixed. The virtuality of the gluon which transfers longitudinal momentum onto the annihilating antiquark scales as $Q^2$. The gluon exchange, as well as
 the subsequent soft rescattering of the stopped quark, are coherent with the photon interaction. The lepton pair senses the entire pion wave function and carries zero helicity in the BB limit.} 
\label{fig5}
\end{figure}

The BB limit \eq{bbl} of the Drell-Yan amplitude is sketched in Fig.~5. As the virtuality $Q^2$ of the lepton pair increases with its momentum fraction $x_F$ such that $(1-x_F)Q^2$ is fixed, the stopped quark with a small fraction $1-x \sim 1-x_F$ remains coherent with the virtual photon interaction. Also the time-scale of the gluon which transfers longitudinal momentum between the quarks in the pion is commensurate with that of the virtual photon. Hence the produced leptons are coherent with the full wave function of the pion. It turns out \cite{Berger:1979du} that the virtual photon is {\it longitudinally} polarized in this kinematic limit, in contrast to the transverse polarization obtained for $Q^2 \to \infty$ at fixed $x_F$. This reflects the conservation of helicity from the pion to the lepton pair.

\begin{figure}[h]
\begin{center}
\psfig{file=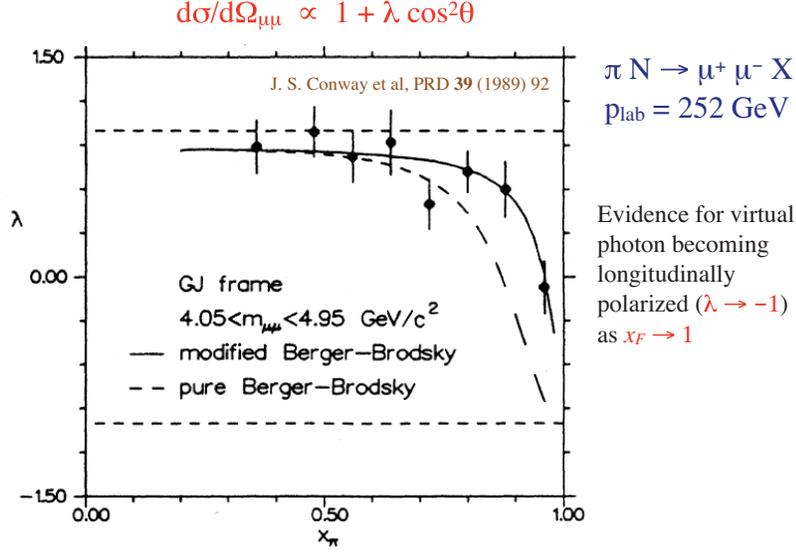,width=11cm}
\end{center}
\caption{E615 data \cite{Conway:1989fs} on the angular distribution of the muon pair in $\pi N \to \mu^+\mu^- X$ at $Q^2 \simeq 20\ \gev^2$. The polarization of the pair changes from transverse towards longitudinal when the pair carries a high momentum fraction of the pion.}
\label{fig6}
\end{figure}

The E615 data \cite{Conway:1989fs} (Fig.~6), while not agreeing with the details of the model calculation, does indeed show a trend for the photon to become longitudinally polarized for $x_F \gsim 0.8$ at $Q^2 \simeq 20\ \gev^2$. It is interesting to note that this effect sets in already at a sizeable $(1-x_F)Q^2 \simeq 4\ \gev^2$. There is evidence \cite{Biino:1987qu} for a similar effect in $J/\psi$ production at high $x_F$.

\subsubsection*{Single spin asymmetry}

In an inclusive process $a+b \to c + X$ parity allows the cross section to depend on the spin component orthogonal to the reaction plane of one of the particles. This single spin asymmetry (SSA) of a spin $\halft$ particle is given by
\beq\label{anexpr}
A_N = \frac{d\sigma^\uparrow-d\sigma^\downarrow}{d\sigma^\uparrow+d\sigma^\downarrow} =
\frac{2\Sigma_{\{\sigma\}} {\rm Im} \left[\M^*_{\leftarrow,\{\sigma\}}\M_{\rightarrow,\{\sigma\}}\right]}{\Sigma_{\{\sigma\}}\left[\left|\M_{\rightarrow,\{\sigma\}}\right|^2+
\left|\M_{\leftarrow,\{\sigma\}}\right|^2\right]}
\eeq
where the $\M$ are helicity amplitudes, and the helicities $\{\sigma\}$ of all particles except the polarized one are summed over. From this expression it is clear that $A_N \neq 0$ requires
\begin{itemize}
\item Helicity flip
\item A dynamical phase (absorptive part)
\end{itemize}
When the produced particle $c$ has large transverse momentum these features must occur in interactions which are {\it coherent} with the hard subprocess for a sizeable asymmetry to occur. However, both helicity flip and absorptive parts are strongly suppressed in perturbative processes, being proportional to current quark masses and $\as$, respectively. Thus it was long ago noted \cite{Kane:1978nd} that $A_N$ should be small in large $p_\perp$ processes.

Data does not agree with these theoretical expectations. The Lambda polarization in $pp \to \Lambda + X$ increases strongly \cite{Lundberg:1989hw} with $x_F(\Lambda)$, reaching values $A_N \gsim 0.3$ at $x_F \simeq 0.8$. $A_N$ furthermore shows no sign of decreasing with transverse momentum, in the measured range $p_\perp(\Lambda) \lsim 3\ \gev$. Similar results (Fig.~7) were later obtained for $p^\uparrow p \to \pi + X$ by E704 \cite{Adams:1991cs} and recently by STAR \cite{Surrow:2007gy}.

\begin{figure}[h]
\begin{center}
\psfig{file=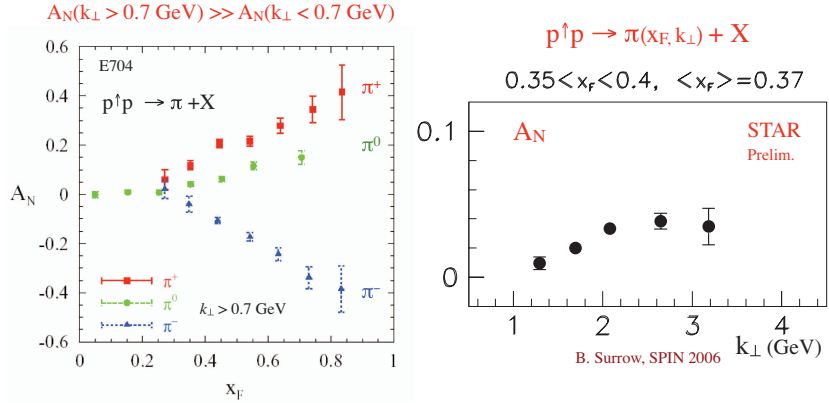,width=11cm}
\end{center}
\caption{The single spin asymmetry \eq{anexpr} in $p^\uparrow p \to \pi + X$. Left: Fermilab E704 data \cite{Adams:1991cs} as a function of $x_F$, for pions with transverse momenta $k_\perp > 0.7$ GeV. Right: The $k_\perp$ dependence from STAR \cite{Surrow:2007gy} at $x_F \simeq 0.37$.}
\label{fig7}
\end{figure}

The fact that $A_N$ does not decrease with transverse momentum up to several GeV is hard to understand in a standard perturbative QCD analysis. $A_N$ is antisymmetric in the transverse momentum $p_{c\perp}$ of particle $c$, and is normalized by the total reaction rate which is even in $p_{c\perp}$. Hence it is inevitable that $A_N \propto 1/p_{c\perp}$ for $p_{c\perp} \to\infty$. It has also formally been shown \cite{Qiu:1991pp} that the SSA is a `twist-3' effect. 

For a single quark to produce a pion with $x_F\simeq 0.8$ (as in the $pp\to \pi+X$ E704 data) it would have to carry $x \gsim 0.9$ of the proton momentum, and then deliver a fraction $z \gsim 0.9$ of its momentum to the pion. At such large values of $x$ and $z$ the quark distribution and fragmentation functions are very small. It was in fact shown \cite{Bourrely:2003bw} that the leading twist cross section is an order of magnitude below the E704 cross section.

A possible way out of this dilemma is that the BB limit \eq{bbl} is more relevant for describing the data than the twist expansion, at the measured values of $x_F$ and $p_\perp$. This is suggested by the increase of $A_N$ for $x_F\to 1$ (Fig.~7) and by the moderate values of $(1-x_F)p_\perp^2$ as compared to the onset of the longitudinal virtual photon polarization in the E615 data on lepton pair production (Fig.~6).

The coherence of soft and hard processes in the BB limit allows the helicity flip and absorptive part required for $A_N\neq 0$ to be generated in a {\it soft part of the amplitude} (Fig.~8), thus avoiding the suppression noted in Ref. \cite{Kane:1978nd}. This was explicitly verified in a perturbative analysis \cite{Hoyer:2006hu}. The large values $A_N \simeq 0.4$ seen experimentally\footnote{The $A_N$ measuresd in DIS is an order of magnitude smaller \cite{Airapetian:2004tw}.} are also easier to understand when the hard process is coherent over the whole wave function of the polarized hadron.

\begin{figure}[h]
\begin{center}
\psfig{file=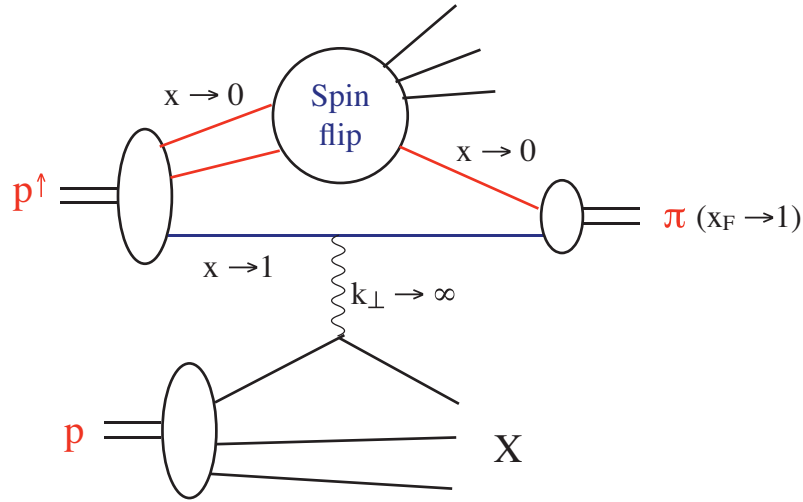,width=11cm}
\end{center}
\caption{Dynamics \cite{Hoyer:2006hu} of the single spin asymmetry in the BB limit \eq{bbl}. The hard scattering occurs off the parton with high momentum fraction $x$ (the energy of the projectile is assumed to much larger than the transverse momentum, $E_p \gg k_\perp$). The soft interactions of the partons with low $x$ are coherent with the hard scattering. Hence a helicity flip and absorptive part in the soft scattering suffices to correlate the pion angular distribution with the projectile spin.}
\label{fig8}
\end{figure}

\subsection*{Bloom-Gilman duality}

Finally I return to Bloom-Gilman duality (Fig.~2), and to what it may teach us about the dynamics of exclusive processes. Consider the target hadron as a superposition of Fock states. The least `miraculous' explanation of duality is that, for a given $x_B$, inclusive and exclusive final states are produced {\it by the same (mixture of) target Fock states}, in the whole range of $Q^2$ at which duality applies \cite{Hoyer:2005nk}. The precocious scaling in inclusive DIS implies that the photon wavelength is small compared to the interparton distance in the Fock state, so that the hard scattering (for exclusive as well as inclusive processes) occurs on a single quark. Coherence between the struck quark and soft partons in the same Fock state is still possible, according to our discussion above of the BB limit \eq{bbl}.

At low values of $Q^2$ the mass $W$ of the hadronic final state is in the resonance region (\cf\ \eq{mrel}). The formation of resonances occurs, however, at a late time, $t_R > L_I$ in \eq{ioffe}, and is thus incoherent with the hard photon scattering. The scattering probability has already been `set', and may only be distributed within the mass uncertainty prevailing at the resonance formation time, $\Delta W \sim 1/t_R$. The semilocal duality of Fig.~2 indicates that $\Delta W$ is of the order of the resonance spacing. Thus unitarity ensures that the resonance bumps average the smooth (high $Q^2$) scaling curve.

\begin{figure}[h]
\begin{center}
\psfig{file=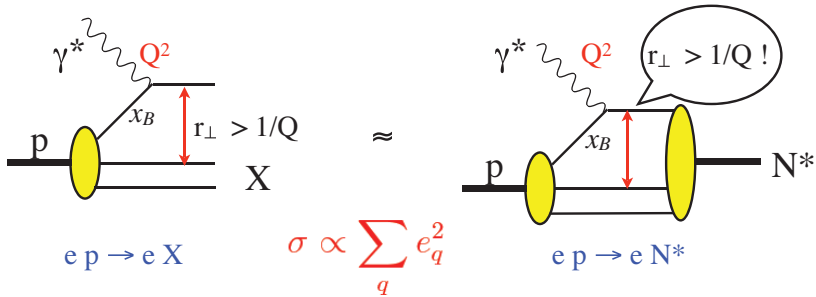,width=11cm}
\end{center}
\caption{Bloom-Gilman duality follows naturally if \cite{Hoyer:2005nk} the virtual photon scatters off the same (superposition of) target Fock states at all $Q^2$. At low $Q^2$ ({\it rhs.}) the formation of the $N^*$ resonance in the final state is incoherent with the photon interaction and thus does not affect the cross section. Scaling at high $Q^2$ ({\it lhs.}) implies that the photon strikes a single quark in the target. Duality then requires this to be the case also for the $p \to N^*$ transition form factor.}
\label{fig9}
\end{figure}

The above scenario requires that target Fock states contributing to exclusive form factors have large transverse size compared to the photon resolution, $r_\perp > 1/Q$. This is the `end-point' contribution in the formalism developed by Brodsky and Lepage (BL) \cite{Lepage:1980fj}. The struck quark carries nearly all the hadron momentum, as in the BB limit \eq{bbl}. While there is no doubt about the BL analysis for compact Fock states with $r_\perp \sim 1/Q$, the importance of the end-points has been debated for a long time \cite{ilr}. The above interpretation of duality suggests that the situation is analogous to DIS, where photon splitting into compact \qq states corresponds to the higher order subprocess $\gamma^* g \to \qu\qbm$, whereas the dominant parton model contribution $\gamma^*\qu \to \qu$ originates from asymmetric states of large transverse extent.

The importance of the endpoint contribution in exclusive processes depends on the behaviour of hadron wave functions in the limit where one parton takes nearly all the momentum, $x\to 1$. It is sometimes assumed that this behaviour may be calculated perturbatively, starting from a non-perturbative wave function with no support at high $x$. However, the E615 data \cite{Conway:1989fs,Melnitchouk:2002gh} indicates that $f_{q/\pi}(x) \propto (1-x)^{1.2}$ for $x\to 1$, disfavouring the perturbative result  $f_{q/\pi}(x) \propto (1-x)^{2}$ for the quark distribution in the pion. Furthermore, in the non-perturbative AdS/CFT approach \cite{Brodsky:2006uq} the pion distribution amplitude $\phi_\pi(x) \sim \sqrt{x(1-x)}$ falls off slower at the endpoints than the perturbative result $\phi_\pi(x) \sim x(1-x)$.

Measurements of color transparency can in principle determine the transverse size of Fock states contributing to, say, $ep \to ep$ at high $Q^2$. In scattering on nuclei, $eA \to ep(A-1)$,  compact proton Fock states emerging from the hard process would be transparent to the nucleus. Present data \cite{Garrow:2001di} shows little evidence for color transparency in these processes, suggesting a large transverse size for the relevant proton Fock states.

\begin{figure}[h]
\begin{center}
\psfig{file=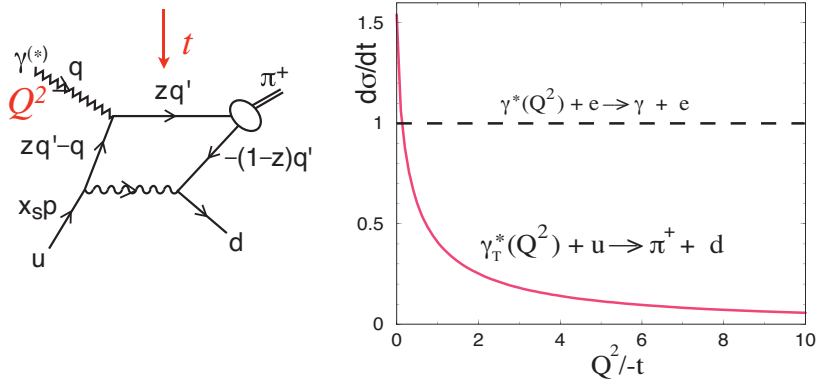,width=11cm}
\end{center}
\caption{The effective size of the QCD subprocess $\gamma +u \to \pi^+ + d$ at large momentum transfer (left) can be `measured' by making the photon slightly virtual \cite{Hoyer:2002qg}. The $Q^2$ derivative of the cross section is logarithmically infinite at $Q^2 =0$ (right), assuming a pion distribution amplitude of asymptotic form. This indicates that the QCD subprocess is not transversally compact, in contrast to QED Compton scattering (dashed line).}
\label{fig10}
\end{figure}

The properties of subprocess amplitudes in QCD allow endpoint contributions to be important. For example, the effective size of the photoproduction subprocess $\gamma u \to \pi^+ d$ at high momentum transfer $t$ is measured \cite{Hoyer:2002qg} by its sensitivity to a small virtuality $Q^2$ of the (transverse) photon (Fig.~10). The derivative of the differential cross section $d\sigma/dt$ {\it wrt.} $Q^2$ turns out to diverge logarithmically at $Q^2 =0$, formally implying an infinite size. The effective size would be even larger for a pion distribution amplitude which vanishes more slowly than the perturbative one at the endpoints. Helicity flip and rescattering effects also enhance endpoint contributions.

\subsection*{Concluding remarks}

Bloom-Gilman duality expresses a remarkable relation between inclusive and exclusive processes. It requires the simultaneous and precocious validity of two distinct high $Q^2$ limits: The Bj limit \eq{bjl} where $x_B$ is held fixed and what we have called the BB limit \eq{bbl}. In the latter limit the mass $W$ of the hadronic system is fixed, and it is thus appropriate for exclusive form factors ($M_{N^*} = W$).

The dynamics of the two limits is very different. All partons in a Fock state where one parton carries nearly all the momentum, $x \sim 1-\lqcd^2/Q^2$, remain coherent in the $Q^2 \to \infty$ limit. I argued that there is experimental evidence of such contributions in the Drell-Yan process and in single spin asymmetries at high $x_F$. This kind of Fock states may also be relevant for exclusive processes, where they appear as endpoint contributions in the standard exclusive framework developed by Brodsky and Lepage.

The multiparton coherence of the BB limit implies that the standard twist expansion is inapplicable. It appears desirable to improve our understanding of this limit, including the factorization of coherent from incoherent scattering dynamics.



\end{document}